\documentclass[12pt]{iopart}

\usepackage{amsmath}
\usepackage[colorlinks=true,     linkcolor=blue, citecolor=blue, filecolor=blue, urlcolor=blue]{hyperref}
\usepackage{graphicx}

\begin{document}
\title[A neural network approach for FT atomic spectra line detection]{A neural network approach for line detection in complex atomic emission spectra measured by high-resolution Fourier transform spectroscopy}

\author{Milan Ding$^1$, Sean Z. J. Lim$^1$, Xiaoran Yu$^2$, Christian P. Clear$^1$, Juliet~C.~Pickering$^1$}

\address{$^1$ Department of Physics, Imperial College London, Prince Consort Road, London SW7~2AZ, UK}
\address{$^2$ Department of Electrical and Electronic Engineering, Imperial College London, South Kensington, London SW7~2AZ, UK}
\ead{milan.ding15@imperial.ac.uk}
\vspace{10pt}
\begin{indented}
\item[]January 2025
\end{indented}

\begin{abstract}
The atomic spectra and structure of the open d- and f-shell elements are extremely complex, where tens of thousands of transitions between fine structure energy levels can be observed as spectral lines across the infrared and UV per species. Energy level quantum properties and transition wavenumbers of these elements underpin almost all spectroscopic plasma diagnostic investigations, with prominent demands from astronomy and fusion research. Despite their importance, these fundamental data are incomplete for many species. A major limitation for the analyses of emission spectra of the open d- and f-shell elements is the amount of time and human resource required to extract transition wavenumbers and intensities from the spectra. Here, the spectral line detection problem is approached by encoding the spectrum point-wise using bidirectional Long Short-Term Memory networks, where transition wavenumber positions are decoded by a fully connected neural network. The model was trained using simulated atomic spectra and evaluated against experimental Fourier transform spectra of Ni ($Z=28$) covering 1800--70\,000~cm$^{-1}$ (5555--143 nm) and Nd ($Z=60$) covering 25\,369--32\,485~cm$^{-1}$ (394--308~nm), measured under a variety of experimental set-ups. Improvements over conventional methods in line detection were evident, particularly for spectral lines that are noisy, blended, and/or distorted by instrumental spectral resolution-limited ringing. In evaluating model performance, a brief energy level analysis of Ni~II using lines newly detected by the neural networks has led to the confident identification of two Ni~II levels, 3d$^8$$(^3\text{F}_4)6\text{f} \,\,[2]_{3/2}$ at 134\,261.8946 $\pm$ 0.0081 cm$^{-1}$ and 3d$^8$$(^3\text{F}_4)6\text{f} \,\,[1]_{3/2}$ at 134\,249.5264 $\pm$ 0.0054 cm$^{-1}$, previously concluded to be unidentifiable using the Ni spectra.
\end{abstract}

%
\vspace{2pc}

\section{Introduction}
Atoms and ions with the most complex spectra contain electrons in the open d- and f-shells \cite{johansson1996term}, where interactions between the multiple valence electrons lead to many thousands or tens of thousands of observable electronic transitions. The experimental determination of fundamental atomic data such as fine structure level energies and angular momenta, and energy level transition wavenumbers is crucial in almost any high to moderate resolution spectroscopic plasma investigations of these species. These data for the open d- and f-shell elements are of interest to a wide variety of fields, such as in the lighting and metal manufacturing industries, in fusion research \cite{von2005complex}, stellar astronomy \cite{atkins2013nailing}, Galactic spectroscopic surveys \cite{heiter2021atomic}, and stellar nucleosynthesis investigations \cite{cowan2021origin}. Additionally, empirical knowledge of atomic energy levels constrains and improves theoretical atomic structure models (e.g. the Cowan codes \cite{cowan1981theory}), offering improvements over \textit{ab initio} estimates. 

High-resolution Fourier transform (FT) spectroscopy is one of the primary experimental methods for the empirical determination of atomic structure of open d- and f- shell species. Recent examples include the open 3d-shell Ni~II \cite{clear2022wavelengths, clear2023wavelengths}, the open 4d-shell Zr~I-II \cite{lawler2022energyzr}, the open 5d-shell Hf~I-II \cite{lawler2022energyhf}, and the open 4f-shell Nd~III \cite{ding2024spectrum1, ding2024spectrum2}. The resolving power of FT spectrometers can be up to a few million, corresponding to wavenumber accuracies that are many orders of magnitude above those achievable with theoretical calculations. Spectral line Doppler and pressure broadening widths of the plasma sources are commonly the factors limiting spectral resolution. With the wide spectral range of FT spectrometers, and with the complexity of atomic structure of the open d- and f-shell species, tens of thousands of transitions between fine structure energy levels of each species are measurable across the mid-infrared (IR) to vacuum-ultraviolet (UV) range.

The immense quantities of transition wavenumbers and intensities of the open d- or f-shell elements are extracted from observed FT spectra into line lists through means of least-squares fitting of model spectral line profiles. Existing methods involve initialising fitting parameters using peak-finding algorithms, optimising the parameters, and then manually including undetected lines or correcting poor fits. This analysis is widely carried out using the FT atomic spectra analysis program Xgremlin \cite{nave2015xgremlin}, with a peak-finding algorithm that averages the peak positions and second derivative peak positions determined using a set of specialised kernels. The greatest challenges arise from unresolved line blending and weak lines that are not discernible from the noise level, causing incorrect line wavenumbers and incomplete extraction of the list of transitions, which hinders the spectrum analyses. Moreover, a relatively high signal-to-noise ratio (S/N) threshold parameter, $smin$, of around 5 is typically set in Xgremlin to avoid noise from being detected as lines \cite{nave2015xgremlin}. The manual adjustments to the line list typically take months for a set of spectra of just one element across the IR-UV. There is a demand for more efficient methodologies.

In addition to dramatically improving the efficiency of the FT atomic spectra line list making process, we also aimed to develop a method to increase the accuracy and completeness of the wavenumbers of weak and blended lines in the line lists. This is important because, the reliable observation of one transition of an energy level could be the key to determine that level and then finding other levels connected through other transitions. As far as we know, there are no records on the application of neural networks (NNs) for line detection in high-resolution FT complex atomic spectra in the literature. In this paper, we present our simplest and highest performing NN model inspired by similar investigations for molecular nuclear magnetic resonance spectra \cite{li2021deep, schmid2023deconvolution}. Our NN model encodes each spectrum point using bidirectional Long Short-Term Memory (LSTM \cite{hochreiter1997lstm}) network layers, and classifies each spectrum point for spectral line detection using a fully connected neural network (FCNN) as the decoder. The LSTM-FCNN model was trained against simulated FT spectra using focal loss \cite{lin2017focal} and evaluated on FT spectra obtained for Ni ($Z=28$) covering 1800--70\,000~cm$^{-1}$ (5555--143 nm) \cite{clear2022wavelengths, clear2023wavelengths} and Nd ($Z=60$) covering 25\,369--32\,485~cm$^{-1}$ (394--308~nm) \cite{ding2024spectrum1, ding2024spectrum2} from a variety of FT spectrometers and atomic emission spectrum light sources. Results show that the LSTM-FCNN model outperforms the peak-finding algorithms of Xgremlin \cite{nave2015xgremlin} in detecting weak lines, blends, and lines affected by instrumental spectral resolution-limited ringing. Around 95\% of the spectral lines in previously human-made line lists were detected by the LSTM-FCNN model. A significant fraction of the NN-detected lines that had not been included previously in the human-made line lists could be confirmed by identification in energy level analyses. As an example, we present the identification of two energy levels of Ni~II that were previously concluded to be unidentifiable from the Ni FT spectra due to incomplete human-made line lists \cite{clear2023wavelengths}.

\section{Methodology overview}
Our approach in using NNs for line detection in FT atomic emission spectra is to formulate the problem as a binary classification task. The LSTM-FCNN models are trained under supervised learning to classify each data point in a FT spectrum, where the class is either a point closest to a transition wavenumber (e.g. line centre, class 1) or not a point closest to a transition wavenumber (class 0). The resulting predictions are then used as initial parameters for spectrum fitting to extract spectral line wavenumber, intensity, and their associated uncertainties for the line list. The LSTM-FCNN models are trained using simulated experimental spectra, primarily because incompleteness and inaccuracies exist in human-made line lists, a problem which we aim for the NNs to overcome. Another motivation for using simulation data is in our objective to reduce labour costs; if the training data costs the same amount of time as it takes to create a human-made line list, the efficiency improvement becomes insignificant. To account for varying spectrum properties between different experiments and elemental species, we simulate spectra and train the NN model independently for each experimental spectrum. 

\begin{figure}
    \centering
    \includegraphics[width=\linewidth]{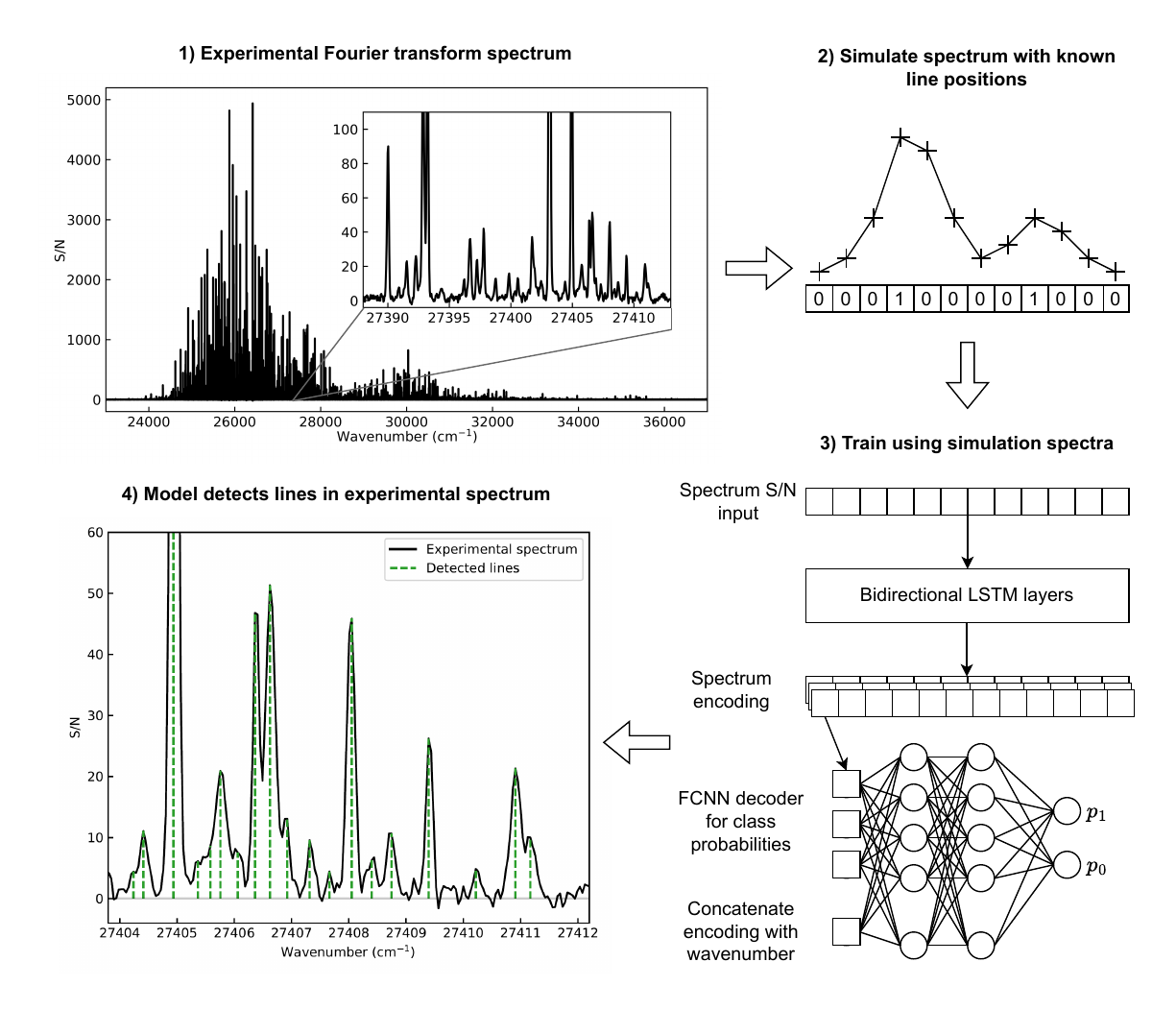}
    \caption{Illustration of the LSTM-FCNN encoder-decoder approach for line detection in a line-dense high-resolution FT atomic emission spectrum within 24\,000--36\,000~cm$^{-1}$ (417--278~nm).}
    \label{fig1}
\end{figure}

We summarise our approach in the illustration shown in Fig.~\ref{fig1}.
The steps are as follows: 1) the experimental spectrum is briefly investigated to estimate simulation parameters such as line width component distributions, number of lines per unit wavenumber, and S/N distributions; 2) a spectrum is simulated with all line positions known; 3) the model is then trained using the simulation data to predict class 1 and 0 probabilities, $p_1$ and $p_0$; and 4) the trained model is applied to detect lines in the experimental spectrum. The remaining sections of this paper detail all four parts of our approach. 

\section{Simulation of high-resolution Fourier transform atomic spectra}\label{sec:sim}
We cannot reliably determine the exact number of spectral lines and their wavenumbers in an experimental spectrum because lines near the noise level or within blends are ambiguous, primarily because the atomic energy levels governing those transitions are commonly unknown. We chose to train the LSTM-FCNN model using simulation spectra because we aim to develop a method to create line lists more efficiently and more completely with fewer undetected lines; we expect training using line lists created by humans using current methodologies would simply not yield efficiency improvements or fewer undetected lines. The simulation of each experimental spectrum is determined by experimental parameters and parameters obtainable from brief investigations of the experimental spectrum. In this section, we discuss the parameters used to simulate the experimental FT spectra used for NN training. All simulation data analysis and calculations are carried out in Python; each simulation of an experimental spectrum takes around a few minutes.

\subsection{General spectrum properties}
A typical high-resolution FT emission spectrum recorded in the laboratory to study the complex atomic structure of an open d- and f-shell species is dominated by transitions from spontaneous de-excitations. The spectrum is an array of length of up to $2^{21}$ (about $2\times10^6$), corresponding to the spectral resolving power at the same order of magnitude. At this resolution, around four data points are sampled across a typical line width, the full width at half maximum (FWHM) of the spectral line. This is a compromise between high wavenumber accuracy (requiring high resolution) and low noise level (to observe weaker lines \cite{davis2001fourier}). The spectral range can be up to a few 10\,000~cm$^{-1}$ (up to a few 100~nm in the visible-UV and up to a few 1000~nm in the IR). The number of spectral lines observed is of order 10$^3$ per spectrum and their observed line profiles are closely approximated by the Voigt profile when fully resolved \cite{thorne1999spectrophysics}. This applies for the two most common discharge sources used for atomic emission FT spectroscopy: the hollow cathode lamp (HCL; e.g. \cite{danzmann1988high}) and the Penning discharge lamp (PDL; e.g. \cite{finley1986penning, heise1994radiometric}). The spectrum is represented as S/N against wavenumber ($\sigma$; in cm$^{-1}$) and has a dynamic range of order 10$^4$.

\subsection{Simulating line widths}
\begin{figure}
    \centering
    \includegraphics[width=\linewidth]{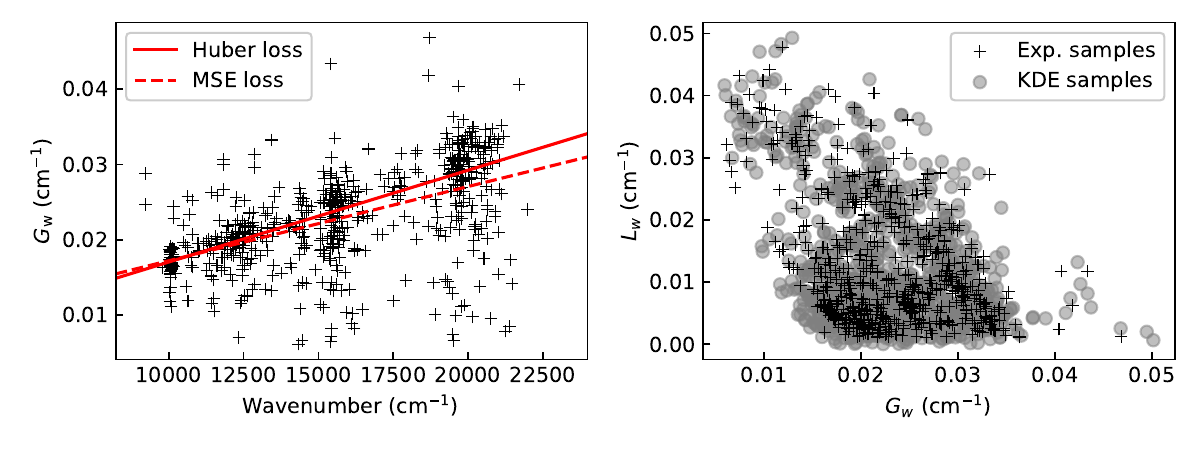}
    \caption{Analysis of relationships between wavenumbers, Doppler widths $G_{\text{w}}$, and pressure widths $L_{\text{w}}$ of spectral lines above 10 S/N in spectrum NiHeDH of the recent Ni~II analysis \cite{clear2022wavelengths}. The left plot shows determination of the linear $G_{\text{w}}$-wavenumber relation using Huber and mean-square-error (MSE) losses, and the y-axis limits are the boundaries of $G_{\text{w}}$ to ensure convergence in spectrum fitting. The right plot shows the correlations between $G_{\text{w}}$ and $L_{\text{w}}$ fitted from experimental spectral lines (crosses) and 1000 $G_{\text{w}}$-$L_{\text{w}}$ pairs sampled using kernel density estimation (KDE; circles).}
    \label{fig2}
\end{figure}

The Doppler broadening Gaussian width ($G_{\text{w}}$) and pressure broadening Lorentzian width ($L_{\text{w}}$) together determine the shapes of normalised Voigt profiles of the experimental spectrum. Generally, $L_{\text{w}}$ is smaller and independent of wavenumber, and $G_{\text{w}}$ is well approximated to depend linearly on wavenumber,
\begin{equation}\label{eq:Gw}
    G_{\text{w}} = \sqrt{\frac{k_{\text{B}}T}{mc^2}}\: \sigma,
\end{equation}
where $T$ is the temperature and $m$ is the mass of the atom or ion. In the present work, the $G_{\text{w}}$ and $L_{\text{w}}$ for a spectrum were obtained by fitting Voigt profiles to spectral lines above 10 S/N, detected by a simple thresholding algorithm that checks for the rise and fall of consecutive data points. Outliers of the $G_{\text{w}}$ and $L_{\text{w}}$ parameters obtained using this method are common and caused by line blending, self-absorption, instrumental artefacts, and line distortions by significant nuclear perturbations such as isotope or hyperfine structure. To account for these outliers, extreme values clustered within 0.001~cm$^{-1}$ of specified boundary values for $G_{\text{w}}$ and $L_{\text{w}}$ were removed. Furthermore, linear fits of the $G_{\text{w}}$-wavenumber relation were carried out using the Huber loss \cite{huber1964robust, scikit-learn}. An example estimation of the $G_{\text{w}}$-wavenumber relation in an IR-visible Ni-He HCL FT spectrum \cite{clear2022wavelengths} is shown in the left plot of Fig.~\ref{fig2},
where outliers exist near the $G_{\text{w}}$ limits, as well as at particular wavenumbers due to instrumental artefacts of the highest S/N lines being detected as spurious lines. The fit using mean-square-error loss was significantly affected by outliers and was poorer compared to the fit using Huber loss. For a spectral line simulated at wavenumber $\sigma$, its $G_{\text{w}}$ is determined using the linear fit for Eq.~\ref{eq:Gw} multiplied by a factor randomly sampled from a normal distribution centered at one and a standard deviation of around 0.3, which is determined from inspecting the spread in $G_{\text{w}}$-wavenumber plots for all experimental spectra. Absolute values are used for the small number of negative samples of the normal distribution centered at one to prevent negative $G_{\text{w}}$ values.

The relationship between $L_{\text{w}}$ and $G_{\text{w}}$ is more complicated. On the right plot of Fig.~\ref{fig2}, clusters of spectral lines are seen. These clusters depend on the plasma light source and approximately correspond to different species in the plasma, such as multiple ionisation stages of the metal and gas elements. Modelling various light source plasmas down to the details of fine structure transition widths is unfeasible, thus we empirically obtain the $G_{\text{w}}$-$L_{\text{w}}$ distribution using kernel density estimation \cite{scikit-learn, rosenblatt1956, parzen1962}. For each simulated $G_{\text{w}}$, 1000 $G_{\text{w}}$-$L_{\text{w}}$ pairs (circles of Fig.~\ref{fig2}) are randomly sampled based on the observed $G_{\text{w}}$-$L_{\text{w}}$ distribution (crosses of Fig.~\ref{fig2}), a corresponding sample of $L_{\text{w}}$ is then randomly selected within $\pm0.001$~cm$^{-1}$ of the simulated value of $G_{\text{w}}$. In rare cases where there are no $L_{\text{w}}$ values within the window, the nearest $L_{\text{w}}$ is selected. Similarly, in rare cases where a $L_{\text{w}}$ sample is negative, the absolute value is used instead.

Lastly, we found that the $G_{\text{w}}$ and $L_{\text{w}}$ distributions of the experimental spectrum should ideally be contained within those of the simulation spectra, because if not then line profiles unseen by the model caused model prediction inaccuracies. For example, when the experimental spectral lines have $L_{\text{w}}$ greater than most $L_{\text{w}}$ of the training data, the model would incorrectly predict weaker lines blended in the wings of each spectral line. This can occur for the relatively small number of spectral lines of the sputtering gas species, with typically lower $G_{\text{w}}$ and higher $L_{\text{w}}$ values, which can in fact guide their classification during spectrum analyses after line list creation.

\subsection{Simulating line positions and line intensities}
Spectral line wavenumber positions and line intensities must also be simulated in addition to simulating their individual $G_{\text{w}}$ and $L_{\text{w}}$ values. Simulated spectral line wavenumbers are uniformly randomly sampled in the spectral range of interest within the experimental spectrum using a specified line density $\rho_{\text{line}}$. The total number of spectral lines randomly sampled is equal to this spectral range multiplied by $\rho_{\text{line}}$. We chose a constant line density independent of wavenumber, which is approximately representative of the spectrum regions of interest with good instrumental response. However, $\rho_{\text{line}}$ is not accurately deducible for an experimental spectrum, because blends and lines near or below the noise level are not reliably detectable. The majority of possible transitions of an open d- or f-shell species are typically low intensity spectral lines, often masked by spectrum noise, and only those transitions with higher branching fractions are observed. The actual line density of a spectrum at a particular intensity therefore increases with decreasing intensity, and approximately follows a power law \cite{learner1982simple}. Hence, simulated line densities and line intensities require joint consideration as we aim to simulate the immense numbers of weak lines. 

In an experimental spectrum, the density of spectral lines with S/N above 10, $\rho_{10}$, is more accurately deducible compared to the total line density $\rho_{\text{line}}$. The value of $\rho_{10}$ is estimated from the spectral line fitting carried out for characterising the $G_{\text{w}}$ and $L_{\text{w}}$ distributions of the experimental spectrum. To simulate the expected large quantities of weak lines in the spectrum with S/N below 10, we linearly extrapolate the histogram of $\log_{10}(\text{S/N})$ obtained for lines above 10 S/N from $\log_{10}(\text{S/N})=1$ to 0, and create a spectrum according to this S/N distribution. 
To ensure model training on the highest S/N lines, at least one line from each of the 10 highest $\log_{10}(\text{S/N})$ histogram bins was sampled. Otherwise, the trained LSTM-FCNN model would ignore lines in the experimental spectra with S/Ns that are much higher than the maximum S/N of the simulated training spectra. We expect $\rho_{10}$ to be a few factors smaller than the total experimental spectrum line density $\rho_{\text{line}}$. In our observations, NN models performed better when trained using simulated spectra that appeared (by inspection) to have slightly higher $\rho_{\text{line}}$ than that of the experimental spectrum, for example, with $\rho_{\text{line}}=10\times\rho_{10}$ the simulated spectrum nearly always resembled the highest line density regions in the experimental spectrum. Such line densities ensure a sufficient number of scenarios with blends and weak lines, while avoiding the introduction of too many lines that would prevent spectrum noise from being learned by the NN models. 

\subsection{Simulating the instrumental line profile}
Each line in an FT spectrum is the convolution of the true line profile of the light source and the instrumental line profile of the spectrometer itself. This instrumental line profile must be simulated correctly to maximise model on experimental spectra. We consider the two largest and most well-defined components of the instrumental line profile, which arise from the finite optical path difference (OPD) and finite aperture size of an FT spectrometer \cite{davis2001fourier}. FT spectra are measured as interferograms in the OPD domain, $x$. The classical FFT algorithms used for transforming most high-resolution FT atomic spectra require that the number of points in the interferogram to be an integer power of two, but the actual number of recorded data points depends on the OPD and sampling rate of the FT spectrometer and is often not a power of two. Therefore, the interferogram of an experimental spectrum is nearly always padded with zeros. We simulate spectra with $N$ that is an integer power of two and multiply the corresponding simulated interferogram by a top hat function centered at zero OPD with width $2L$ equal to the experimental maximum OPD. We also include the finite aperture effect through multiplication of a sinc function with the interferogram, where $\text{sinc}(y)=\sin(y)/y$. Thus, the instrumental line profile is added to each simulated Voigt profile spectral line at wavenumber $\sigma$ by multiplying the interferogram of the simulated line with the apodising function
\begin{equation}\label{eq:instrumental_apodisation}
A(x) =
\begin{cases} 
\text{sinc}(\sigma\Omega x / 2) & \text{if }  -L \leq x \leq +L, \\
0 & \text{otherwise},
\end{cases}
\end{equation}
where $\Omega = \pi / L\sigma_{\text{max}}$ is the solid angle subtended by the spectrometer aperture as viewed from the centre of the spectrometer collimating mirror, and $\sigma_{\text{max}}$ is the upper wavenumber limit of the spectrum. The relation for $\Omega$ assumes the standard experimental practice of choosing an aperture size that maximises the amplitude of the sine function (in the numerator of the sinc function) at $\pm L$ for $\sigma_{\text{max}}$. Additionally, resolution-limited ringing of the instrumental profile from the discontinuity at $\pm L$, due to the FT of a top hat function, is reduced in practice by smoothly bringing the interferogram down to zero at $\pm L$ \cite{davis2001fourier}. We apply a cosine bell such that the apodisation function $A(x)$ is further multiplied by $1 + \cos{\big([x \pm (L-aL)]\pi/aL\big)}$ for $\mp L < x < \mp L \pm aL$, where $0 < a \leq 1$. The parameter $a$ is 0.05 by default in Xgremlin \cite{nave2015xgremlin}, and is typically increased in under-resolved spectra to reduce resolution-limited ringing distortions. The S/Ns of the simulated line before and after convolution with the instrumental line profile are ensured to be the same using the ratio of peak values before and after convolution. Computational costs are reduced by applying the apodisation to the FT of a window of spectrum points centered around each simulated line, as opposed to the FT of the entire wavenumber domain with potentially millions of points. The width of this window is chosen such that it contains all resolution-limited ringing elements in the spectrum that are significantly above the noise level. The instrumental profile is convolved with all simulated spectral lines above S/N of 2.

\subsection{Consideration of noise and other instrumental artefacts}
The noise level defines the S/N axis of each spectrum, its appearance also vary between experimental spectra. Generally, noise in FT spectra arises from many sources such as photon noise, digitising noise, light source noise, and background noise. Together, these amount to a characterisable white noise \cite{davis2001fourier}. After standardisation, the histogram of spectrum point S/N values without spectral lines (the white noise of the spectrum) closely follows a normal distribution with Kolmogorov-Smirnov test $p$-value of around $0.75$. Thus, we simulate noise in FT spectra using white Gaussian noise of standard deviation equal to one as the spectra are considered on the S/N scale. 

Deviations from white noise are however common. In the neighbourhood of very high S/N spectral lines, discrepancies from the ideal Voigt and instrumental (Eq. \ref{eq:instrumental_apodisation}) line profiles and phase errors from the FT \cite{davis2001fourier, learner1995phase} become particularly significant. In the infrared spectral region, inaccurate removal of the thermal background and unwanted molecular absorption features also add another layer of complexity to instrumental artefacts. Therefore, deviations from the flat white noise arise from both systematic and random effects, which are not well-defined and often occur at wavenumber scales similar to spectral line widths. We partially address this problem by applying a wavenumber dependent S/N scaling for the experimental spectra with significant deviation from white noise, this is discussed in more detail in section \ref{sec:preprocessing} covering spectrum preprocessing.

\subsection{Creating the binary classification target values}
For our supervised learning binary classification approach, we must systematically decide which points in a simulated spectrum are closest to the line centres, (class $1$) and which are not (class $0$). Inevitably, the simulated spectral lines with randomly generated positions, S/Ns, and widths must be separated into detectable and non-detectable groups. For example, a line with S/N of 0.1 is unlikely to be detectable, and the same applies for a line with S/N of 10 that is masked by a blend with a line with S/N of 10$^3$, because the simulated Voigt profiles are not accurate to experimental line profiles to more than one part in 10$^3$ (e.g. deviations contributed by random phase errors, self-absorption, isotopic shifts, and hyperfine structure). Regardless of the decisions on which simulated lines are detectable, they are all still added to the simulated spectrum to mimic experimental spectra. However, only those lines classed as detectable contribute to class $1$ of the target array used for model training.

Similar to experimental spectra, simulated spectral lines exist below the noise level, and so a threshold S/N$_{\text{min}}$ is used to decide whether a sampled line is detectable. However, lines with S/Ns below S/N$_{\text{min}}$ could together, in a blend, produce a spectral feature above S/N$_{\text{min}}$ depending on their wavenumber separation. We address this by combining lines sampled below S/N of 5 with wavenumbers within four times the spectral resolution ($1/2L$) before any of the randomly generated lines are added to the simulated spectrum. These two thresholds were chosen by considering that four points typically lie between the FWHM and such blends are unlikely to be detectable below S/N of 5. 

Decisions on whether lines are masked and undetectable due to blending with much stronger lines or resolution-limited ringing are challenging as they depend on the separation between lines and complex line profiles. As each simulated spectral line is added to the simulated spectrum one by one, after convolution with its wavenumber-dependent instrumental profile, this allows comparison between the spectrum simulated so far and the line to be added to decide whether the line to be added is detectable. We add simulated spectral lines to the spectrum in decreasing S/N order, as the strongest lines of each blend are likely to be the most detectable and unlikely be masked. For each simulated spectral line wavenumber, the closest four points in the spectrum simulated so far are compared with its simulated S/N; if the simulated S/N is less than 20\% of the highest S/N of the absolute value of the four neighbouring points in the spectrum, it is considered undetectable. The 20\% threshold was decided from trial and error evaluations of model performance on experimental spectra. After all simulated spectral lines have been added to the simulation spectrum, all line wavenumbers classed as detectable but unresolvable (lie closer to another line than the experimental spectral resolution $1/2L$) are combined with their neighbour by S/N weighted average to yield the final list of wavenumbers used to assign the target values for supervised learning. 


\section{Applying the neural network model} \label{sec:model}
Additional steps are required in preparing simulation and experimental data for the LSTM-FCNN model training and evaluation, as well as in using the output class probabilities $p_1$ and $p_0$ to create a line list for each experimental spectrum. In this section, we discuss these data processing aspects of our approach and the training of the LSTM-FCNN model.

\subsection{Spectrum preprocessing} \label{sec:preprocessing}
High-resolution FT atomic emission spectra can be millions of points in length. Wavenumber dependence of general line profile properties, e.g. $G_{\text{w}}$ and $A(x)$, vary smoothly across a spectrum, so spectrum chunking is applicable in reducing memory costs. However, due to the large number of lines and high line densities, a fraction of spectral lines lying on chunk edges will not be fully covered within a chunk. This issue was overcome by using overlapping windows when chunking the experimental spectra, where only model predictions in non-overlapping regions are taken and concatenated back together. The overlapping region was one quarter of the chunk width from both edges. In training, the simulated spectra are chunked without overlapping windows to avoid bias in overlapping regions. The chunk width $N_{\text{chunk}}$ was chosen such that the local noise and any simulated resolution-limited ringing are sufficiently sampled. The typical chunk width is about 50~cm$^{-1}$. 

\begin{figure}
    \centering
    \includegraphics[width=0.8\linewidth]{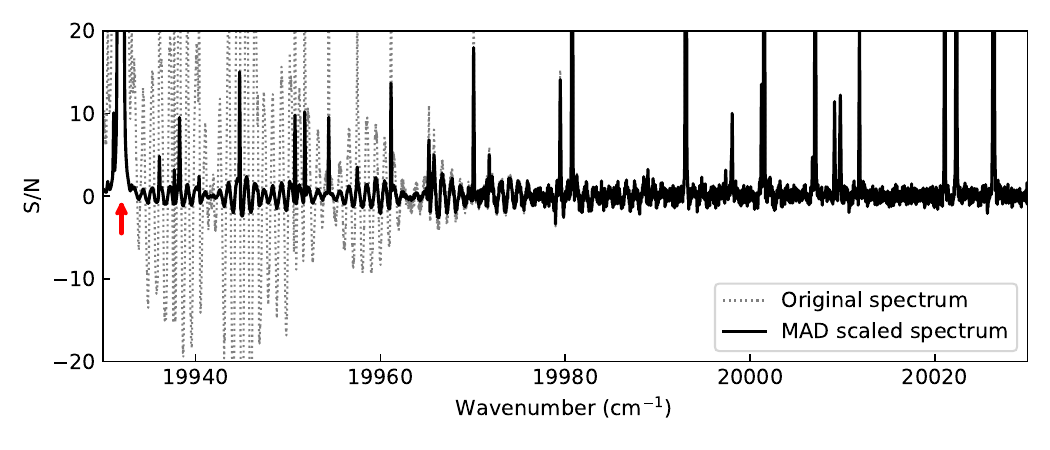}
    \caption{A Ni-He HCL FT spectrum section with (solid) and without (dotted) noise scaling. The artefacts from the 20\,000~S/N He line at 19932~cm$^{-1}$ (indicated by the arrow) on the left are rescaled to match the noise level seen in the remainder of the spectrum, for example the typical noise level seen on the right of the plot.}
    \label{fig3}
\end{figure}

When noise in the experimental spectra is significantly larger than, and deviates from, the simulated white Gaussian noise due to uncharacterisable instrumental artefacts and noise, the noise level of each chunk was re-scaled. Each experimental spectrum chunk was scaled such that the median absolute deviation (MAD) of the S/Ns of the chunk is equal to 0.6745 (the MAD value of Gaussian white noise). The MAD quantity of the S/Ns of a chunk is insensitive to the relatively few points falling on spectral lines and offers a convenient estimate of the new noise level used to match better with the simulation data used for model training. This scaling reduced spurious lines being detected and resulted in more reliable detections near strong lines. An example is shown in Fig.~\ref{fig3},
where the oscillatory ringing artefacts of the He ($Z=2$) line in the original spectrum were scaled to the noise level, which would otherwise be detected as a dense group of false lines. The artefacts of this He line reached above 50 in S/N and appeared to contain multiple frequency components, and their decay was not well-defined. These artefacts cannot be reliably removed by isolating their frequency components in the interferogram because of the overwhelming signal from the 20\,000~S/N He line. Similarly, background approximation methods are also unsuitable as the oscillatory periods of the ringing are comparable to spectral line widths. When MAD scaling is applied, the experimental S/N and width distributions are determined using the MAD scaled spectrum. For spectra with immense line densities, such as those of the lanthanides, the MAD scaling method will overestimate the noise level and is not recommended as there are seldom points in the spectrum without spectral lines, in which case a more detailed wavenumber-dependent noise level scaling may be required.

Spectrum interpolation can slightly improve accuracy of predicted line positions, especially for strong lines with no signs of blending and for spectra with interferograms that only required a small amount of padding to reach an integer power of two in length (thus with only a small amount of interpolation already applied). However, this improvement plateaus as more than one point is interpolated in the spectrum interval, especially in the widespread line blending scenarios when uncertainties in component line positions are comparable with the spectral resolution. A compromise between accuracy improvement and simulation plus training time must be met. In this work, we either applied no interpolation to the experimental spectra or increased the number of points by a factor of two through interferogram padding. After preprocessing, the simulated spectra and experimental spectra are input into the model for training and evaluation.

\subsection{Model architecture}
\begin{figure}
    \centering
    \includegraphics[width=0.9\linewidth]{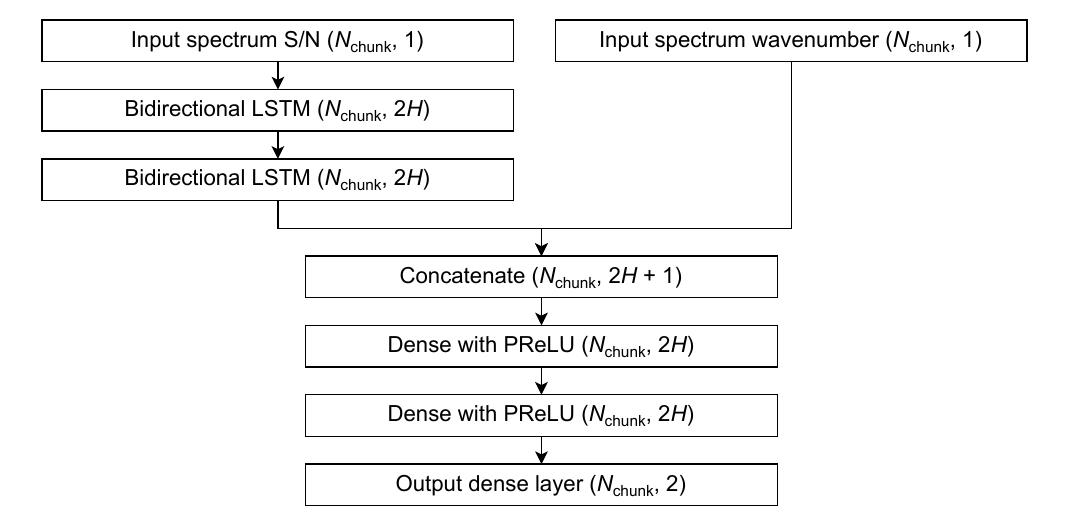}
    \caption{Layout of the bidirectional LSTM-FCNN encoder-decoder model.}
    \label{fig4}
\end{figure}

Our model is constructed using the PyTorch package \cite{NEURIPS2019_9015}. We apply two layers of bidirectional LSTMs to encode the input spectrum chunk and a fully connected NN with two hidden layers to decode output of the LSTMs for probabilities of class $1$ and $0$ for each point in the chunk. The LSTMs are capable of learning individual and blended spectral line profiles, as well as the regions of noise in spectrum chunks that are crucial to compare with when identifying weak lines. The output of the bidirectional LSTM layers are also concatenated with the wavenumber of the spectrum point to include the wavenumber dependence of $G_{\text{w}}$ and Eq. \ref{eq:instrumental_apodisation}. The wavenumber is normalised within the spectral range of interest of the entire spectrum to values between 0 and 1. Lastly, ReLU activation functions for the FCNN layers were found to perform poorly compared to activation functions with non-zero response to negative input. 
For efficiency and adaptability, PReLU \cite{he2015prelu} was selected. Figure~\ref{fig4} shows the layout of the model architecture, where $H$ is the number of hidden states in the bidirectional LSTM layers. 
An $H$ of $64$ was identified as a sensible compromise between performance and training time, resulting in a total number of parameters of about 170\,000.

\subsection{Model training}
A model is trained for each experimental spectrum. We applied a train-test split ratio of 80-20 to chunks of the simulated spectra. Adam \cite{kingma2015adam} is used to optimise model parameters with learning rate at $0.001$ and a batch size of 256 chunks. The loss function is the sum of focal loss \cite{lin2017focal} from each point in each chunk of the batch,
\begin{equation}
    L = -\alpha_{\text{t}}(1-p_{\text{t}})^2 \log(p_{\text{t}}),
\end{equation}
where $\alpha_{\text{t}}$ is the class imbalance weight determined from ratios between simulated classes $1$ and $0$ in the training dataset, and $p_{\text{t}}$ is the predicted probability of the true class. Compared to weighted cross-entropy, focal loss weighs difficult-to-predict spectrum points significantly more, reduced training time, enabled more precise classified line positions, and improved detection within blends. To avoid focus on chunk edges that may contain incompletely sampled lines that are also difficult for the LSTM-FCNN model to predict, the loss contribution is masked at zero near the chunk edges.

\begin{figure}
    \centering
    \includegraphics[width=0.8\linewidth]{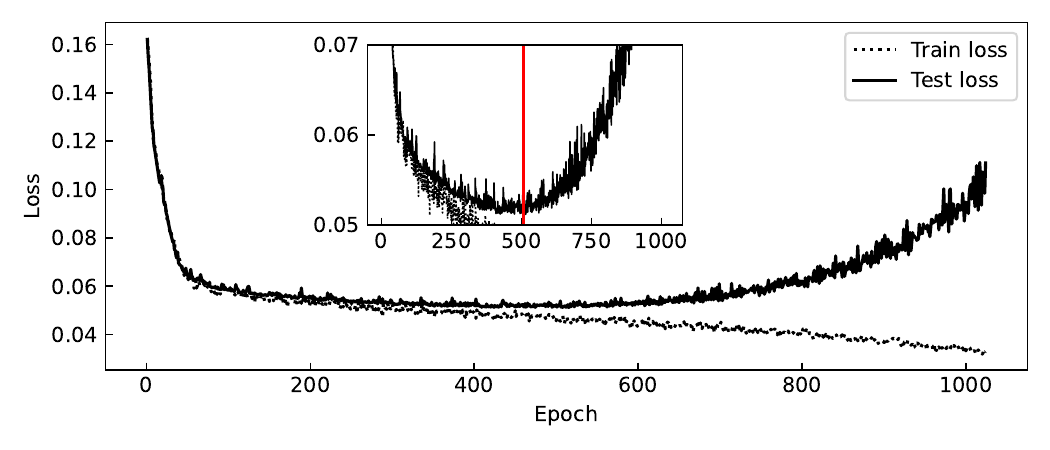}
    \caption{Example loss from the training dataset (dotted) and test dataset (solid) over 1024 epochs during training for simulated spectra with $\rho_{\text{line}}\sim2$ per cm$^{-1}$. The minimum test loss is indicated by the vertical line in the y-axis expanded plot.}
    \label{fig5}
\end{figure}

Model performance on the test dataset, which was comprised entirely of simulation spectra chunks, is only partially indicative of the eventual model performance on the experimental spectrum. However, the change in train and test losses with the number of training epochs is still useful in choosing hyperparameters such as the total number of epochs, number of hidden states $H$, line density $\rho_{\text{line}}$, and distributions for the widths $G_{\text{w}}$ and $L_{\text{w}}$. Specifically, overfitting can occur at different number of epochs depending on data and model complexity, an example is shown in Fig.~\ref{fig5} for a model with $H=64$ and trained using spectra simulated for the high line density Nd-Ar PDL FT spectrum.
With increasing number of training epochs, loss from the training data decreased indefinitely and did not appear to plateau, while the loss for the test data started increasing after some number of epochs, indicating overfitting. Typically, a higher line density required a higher number of epochs to reach overfitting due to increasing data complexity. The trained model is chosen as the model state with the lowest test loss, for the example in Fig.~\ref{fig5} this was the model state after epoch number 507. Training a model with $H=64$ for 1024 epochs using one simulated spectrum may cost up to around 5 minutes on a GPU for spectrum length $2^{21}$ (about $2\times10^6$) with early stopping and adaptive learning rate. 

Prediction accuracy on test simulation spectra is not a useful metric as the number of class 0 points overwhelms the number of class 1 points. Since we are most interested in detecting as many lines as possible, the recall and precision metrics which focus on true positives provided better insights for model performance on training data. Across the line density range observed for Ni ($\rho_{\text{line}}\sim0.2$ per cm$^{-1}$) and Nd ($\rho_{\text{line}}\sim2$ per cm$^{-1}$) spectra, recall decreased from 0.99 down to 0.92 with increasing line density at threshold probability $p_{\text{th}}=0.5$ for classification, implying at least 90\% of the simulated training spectral lines were detected at this threshold. The decrease in recall with increasing $\rho_{\text{line}}$ was expected due to the increased number of weak lines near S/N$_{\text{min}}$ and masked in blends. The precision score is partially representative of the uncertainties of model predictions on the exact locations of the detected lines, because the false positives almost always occurred adjacent to true positives. At $p_{\text{th}}=0.5$, the precision score varied between 0.1 and 0.5. This implied that on average at least two points were predicted with $p_1>0.5$ per detected line. The variation in precision was mainly dependent on the class imbalance contributed by $\rho_{\text{line}}$ and also the degree of interpolation of the spectrum, as the number of class 0 points across a line increases with interpolation, while there is only one class 1 point regardless of interpolation. The low precision scores were not problematic for our purposes as the highest $p_1$ position within each spectral line generally corresponded very well with the peak position.

Model performance on experimental spectra is challenging to define, because we do not know the exact locations of the spectral lines and cannot detect all of them reliably. Therefore, we investigated performance of the model with the lowest test loss on experimental data. Results from these analyses are detailed in section \ref{sec:eval}.


\subsection{Model predictions and postprocessing}
\begin{figure}
    \centering
    \includegraphics[width=\linewidth]{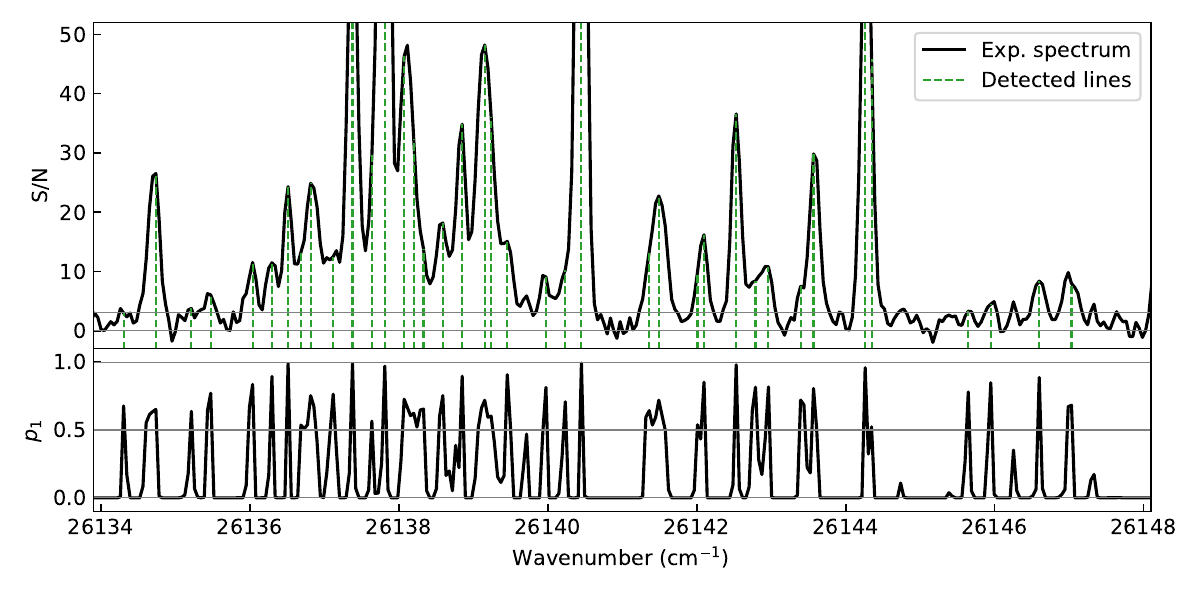}
    \caption{Section of the experimental Nd-Ar PDL FT spectrum with the highest line density (top) and class 1 probability curve of the trained LSTM-FCNN output (bottom). On the spectrum plot, the vertical dashed lines show spectral lines detected from postprocessing and the solid horizontal lines show zero S/N and the S/N$_{\text{min}}=3$ threshold used for training spectra simulations. Horizontal lines on the probability plot shows $p_1=0,0.5,\text{and}\:1$, where $p_1=0.5$ was the threshold for line detection.}
    \label{fig6}
\end{figure}

As discussed previously in section \ref{sec:preprocessing}, class probabilities in non-overlapping regions of neighbouring chunks of the experimental spectra are concatenated. To retrieve line positions from the predicted class probability curves across a spectrum, we apply a threshold probability $p_{\text{th}}$ of 0.5 on the probability curve of class 1, $p_1$, to first extract groups of consecutive class 1 wavenumber positions. Then, single or multiple line positions are selected by a peak-finding algorithm which compares the rise and fall of consecutive $p_1$ values within each group. An example result of using this method is shown in Fig.~\ref{fig6}.
The probability curve $p_1$ resembles a de-noised higher resolution spectrum. 
Typically, lines nearer to the $p_1=0.5$ threshold are those near the S/N$_{\text{min}}=3$ threshold and/or those that are the weaker lines of a blended line profile. Furthermore, the width of the peaks in the $p_1$ curve increases for lines showing anomalously large widths and blended features. Therefore, the $p_1$ curve will be a further aid to human decision making when manual adjustments to the spectrum fitting and line list are required.


The S/Ns of the identified line positions are indicative of line intensity, and were used as initial parameters for Voigt profile fitting of the entire spectrum for the output line list of wavenumbers, relative intensities, and their uncertainties. As is evident from Fig.~\ref{fig6}, our approach with the LSTM-FCNN model for line detection does not completely eliminate manual adjustments. Firstly, a few weak lines could arguably be added to the LSTM-FCNN detections. This could be addressed by lowering S/N$_{\text{min}}$ or $p_{\text{th}}$, which may induce more false detections in the noise and thus still require manual adjustments. Secondly, in widespread line blending scenarios, multi-Voigt profile fits converge poorly and thus human inspection is required after spectrum fitting. Lastly, unresolved but potentially detected isotope or hyperfine structure component lines may require S/N weighted wavenumber averaging, commonly referred to as centre-of-gravity (COG) fitting, for consistent measurements of fine structure energy level separations. More details on whether weak and blended lines in experimental spectra are correctly detected by the LSTM-FCNN model will be discussed in the next section.

\section{Model evaluation on experimental Fourier transform atomic spectra}\label{sec:eval}
To investigate the performance of the LSTM-FCNN model on experimental spectra, we created line lists using our LSTM-FCNN models trained for each of the nine Ni-He HCL FT spectra covering 1800--70\,000~cm$^{-1}$ (5555--143 nm) used for the atomic structure analysis of Ni II \cite{clear2022wavelengths,clear2023wavelengths}. The Ni-He HCL FT spectra were recorded using three different instruments, the VUV FT spectrometer at Imperial College, the 2 m FT spectrometer at the National Institute of Standards and Technology (US), and the 1 m FT spectrometer at Kitt-Peak National Solar Observatory (US). These Ni-He HCL FT spectra were used to evaluate our LSTM-FCNN model approach on spectra from different experimental set-ups and its ability to correctly detect spectral lines that were missed by laborious but careful human analyses. Furthermore, to investigate performance on the highest line density atomic FT spectra, we also trained the LSTM-FCNN model to create a line list for the highest line density Nd-Ar PDL FT spectrum used for the analysis of Nd III covering 25\,369--32\,485~cm$^{-1}$ (308--394~nm) \cite{ding2024spectrum1}. This section summarises our findings on applying our model to real experimental spectra to create line lists of wavenumbers and intensities.

\subsection{Overall comparisons with human-made line lists}
We investigated the overlaps between: 1) the NN line list created over the course of a few hours using Voigt profile fitting of the LSTM-FCNN model line detections, 2) the human-made line list created over the course of a few months from applying peak-finding algorithms of Xgremlin, Voigt profile fitting, and manual adjustments, and 3) the list of Ritz wavenumbers\footnote{The Ritz wavenumber of a fine structure transition is the difference between its level energies that have been optimised using measured transition wavenumbers between all known levels of the species.} $\sigma_{\text{Ritz}}$ of allowed electric dipole transitions between experimentally known atomic energy levels. Note that all numbers of lines quoted above 1000 are rounded to the nearest 100 for clarity, as the number of lines in a line list is rarely accurate to the nearest 10.

For the nine Ni-He HCL FT spectra covering 1800--70\,000~cm$^{-1}$ (5555--143 nm) \cite{clear2022wavelengths, clear2023wavelengths}, 13\,400 lines were detected by the NN approach in total. The key parameters used for spectrum simulation, model training, and data processing are summarised in Table~\ref{tab:Ni-He_params}. 
\begin{table}
\caption{\label{tab:Ni-He_params}Spectrum simulation, model training, and data processing parameters used for each of the nine Ni-He HCL FT spectra.}
\footnotesize
\begin{tabular}{@{}ccccccccc}
\br
$\rho_{\text{line}}$ & S/N$_{\text{min}}$ &  $N_{\text{sim}}$$^{a}$ & $N_{\text{chunk}}$ & $H$ & $N_{\text{epoch}}$$^{b}$ & MAD scaling?$^{c}$ & $p_{\text{th}}$$^{d}$ \\
\mr
$10\times\rho_{10}$ & 3 & 4 & 1024 & 64 & 256 & Yes & 0.5 \\
\br
\end{tabular}\\
$^{a}$number of spectra simulated and used for model training; $^{b}$maximum number of epochs in training; $^{c}$whether chunk noise level was rescaled using median absolute deviation; $^{d}$threshold probability for class 1.
\end{table}
\normalsize
In contrast, the number of lines in the human-made line list is 6700 \cite{clear2022wavelengths, clear2023wavelengths}, approximately half of the number of lines detected by the NN. Using a wavenumber matching tolerance of 0.05~cm$^{-1}$ (typical largest experimental wavenumber uncertainty) to check lines common to both lists, 96\% of the lines in the human line list were detected in the NN line list. The remaining 4\% of the human line list consisted of weak lines and COG wavenumbers. Disagreement between COG fitted wavenumbers is expected as the LSTM-FCNN models were only trained using blended Voigt profiles and have no experience with spectral lines that exhibit significant isotope structure for which COG fitting is carried out. The other half of the 13\,400 lines that were not present the human line list were mostly weak lines, not very obvious blends, isotope component lines, and instrumental artefacts. Of these 6700 lines, 1900 matched with known Ritz wavenumbers of allowed electric dipole transitions of Ni II (excluding Ni I) using the 0.05~cm$^{-1}$ tolerance. A brief comparison between the estimated theoretical and observed relative intensities of the 1900 lines indicated that many of them could be chance coincidences, because of insufficient level populations and transition probabilities. We cannot determine how many of these 1900 lines are real because the known list of Ritz wavenumbers of allowed transitions is incomplete (due to unknown energy levels). However, a non-negligible fraction of these 1900 lines did indeed agree with their expected relative intensities from the theoretical calculations \cite{clear2022wavelengths, clear2023wavelengths, kurucz2017including}. The correct detection of weak lines missed by humans was then additionally supported by a brief extension of the atomic structure analysis of Ni~II (more details in \ref{sec:results_weak_lines}).

For the single Nd-Ar PDL FT spectrum with the highest line density, covering 25\,369--32\,485~cm$^{-1}$ (308--394~nm), 9400 lines were detected by the LSTM-FCNN model trained under the same parameters of Table~\ref{tab:Ni-He_params}, but with $N_{\text{epoch}}=1024$ and without MAD noise scaling due to negligible instrumental artefacts and high line densities. The number of lines in the human line list for this spectrum was 7000 \cite{ding2024spectrum1}, of which 6600 (94\%) were matched with the NN line list under a tolerance of 0.05~cm$^{-1}$. Similarly to the case for Ni, the 400 human detected lines not matched with any lines from the NN line list were from differing detections for blended and weak lines, and lines showing significant isotope shifts. A more detailed investigation on blended line detection by the LSTM-FCNN model will be discussed in section~\ref{sec:results_blended_lines}. Unlike the case for Ni, the increase in the number of lines compared to the human-made line list is relatively smaller, primarily because there were no uncharacterisable instrumental artefacts in this UV Nd spectrum contributing to potential false detections. Matching the NN line list of the Nd-Ar PDL FT spectrum with Ritz wavenumbers of Nd~I-III was not performed because, compared to Ni~II, the knowledge of energy levels of these species is much more incomplete and uncertainties of most known level energies are also larger \cite{blaise1971present, blaise1984revised}, and with the much higher line densities in the Nd-Ar PDL FT spectrum, far more coincidental wavenumber matches would be induced and provide less meaningful estimates. 

\subsection{Weak lines and newly confirmed Ni II energy levels}\label{sec:results_weak_lines}
We carried out a focussed energy level analysis of Ni~II using the NN linelist created from the nine Ni-He HCL FT spectra and calibrated to the same wavenumber and intensity scales as those of the human-made line list \cite{clear2022wavelengths}. We searched for several energy levels that could not be reliably identified using the human-made line lists created for these spectra; their lines were concluded to be too weak \cite{clear2022wavelengths,clear2023wavelengths} and some of which were not present in the human-made line list. The addition of weak lines newly detected by the LSTM-FCNN model increased confidence in the identification of many of these levels. Furthermore, the identification of several levels were concluded to be only possible using the NN line list, two of which are presented in Table~\ref{tab:ni2_lines}.
\begin{table}
\caption{\label{tab:ni2_lines}Classified transitions for two newly optimised energy levels of Ni II.}
\footnotesize
\begin{tabular}{llrllll}
\br
Level$^{a}$  & Line list$^{b}$ & S/N & \multicolumn{1}{c}{$\sigma$}      & $\sigma$ unc.   & \multicolumn{1}{c}{$\sigma-\sigma_{\text{Ritz}}$$^{c}$}     & Lower level$^{d}$ \\
       &        &       & \multicolumn{1}{c}{(cm$^{-1}$)}   &  (cm$^{-1}$)    & (cm$^{-1}$)    &       \\
\mr
\multicolumn{5}{l}{3d$^8$$(^3\text{F}_4)6\text{f} \,\,[2]_{3/2}$ at 134\,261.8946 $\pm$ 0.0081 cm$^{-1}$} \\ 
   & Human         & 9   & \,\,\,\,4977.4780        & 0.0120     & -0.0040       & 3d$^8$$(^3\text{F})6\text{d} \,\,^4\text{P}_{5/2}$ \\
   & Human         & 5   & 14\,596.7015       & 0.0084     & \,\,0.0026        & 3d$^8$$(^3\text{F})5\text{d} \,\,^4\text{P}_{5/2}$ \\ 
   & NN            & 3   & 35\,700.8209       & 0.0320     & -0.0120       & 3d$^8$$(^3\text{F})4\text{d} \,\,^4\text{P}_{5/2}$ \\ 
   \\
\multicolumn{5}{l}{3d$^8$$(^3\text{F}_4)6\text{f} \,\,[1]_{3/2}$ at 134\,249.5264 $\pm$ 0.0054 cm$^{-1}$} \\
   & Human        & 12  & \,\,\,\,4407.3141        & 0.0023     & \,\,0.0008        & 3d$^8$$(^3\text{F})6\text{d} \,\,^2\text{D}_{5/2}$ \\ 
   & NN           & 8   & \,\,\,\,4769.9564        & 0.0077     & \,\,0.0137        & 3d$^8$$(^3\text{F})6\text{d} \,\,^2\text{P}_{3/2}$$^{e}$ \\
   & Human        & 6   & 14\,539.9760       & 0.0130     & -0.0140       & 3d$^8$$(^3\text{F})5\text{d} \,\,^2\text{P}_{3/2}$ \\
   & NN           & 3   & 34\,690.3337       & 0.0330     & -0.0280       & 3d$^8$$(^3\text{F})4\text{d} \,\,^4\text{D}_{5/2}$ \\ 
   & NN           & 3   & 35\,208.9502       & 0.0250     & -0.0250       & 3d$^8$$(^3\text{F})4\text{d} \,\,^2\text{P}_{3/2}$ \\ 
\br
\end{tabular}\\
$^{a}$energy optimised using LOPT \cite{kramida2011program} following procedures from the previous Ni~II energy level analysis \cite{clear2022wavelengths, clear2023wavelengths};
$^{b}$line list from which the wavenumber and its uncertainty were taken from, all four lines in the human-made line list were also present in the NN line list, but none of the four lines in the NN line lists were present in the human-made line list;
$^{c}$difference between observed and Ritz wavenumbers;
$^{d}$combining level designations from the previous Ni~II energy level analysis \cite{clear2022wavelengths, clear2023wavelengths};
$^{e}$line located in a region affected by ringing from a nearby strong line affecting observed line position, this line was excluded from the optimisation of the level energy but is included in this table for reference.
\end{table}
\normalsize
The new 3d$^8$$(^3\text{F}_4)6\text{f} \,\,[2]_{3/2}$ level now found at 134\,261.8946 $\pm$ 0.0081 cm$^{-1}$ was previously unknown. The 3d$^8$$(^3\text{F}_4)6\text{f} \,\,[1]_{3/2}$ level at 134\,249.5264 $\pm$ 0.0054 cm$^{-1}$ was previously proposed by lower wavelength precision grating spectroscopy analysis 50 years ago \cite{shenstone1970second}, but had been unconfirmed in the recent Ni-He HCL FT spectrum term analysis \cite{clear2022wavelengths}. Without the four lines newly detected by the LSTM-FCNN model, identifying each of these two levels had involved looking for a pair of lines at the expected relative intensity ratio, with wavenumbers corresponding to the energy separations between the level and two known levels. This was not possible as there were too many ambiguous pairs of lines matching this description. However, with the additional lines newly detected by the LSTM-FCNN model, the increase in the number of observed lines for each level fully removed such ambiguities.

\begin{figure}
    \centering
    \includegraphics[width=\linewidth]{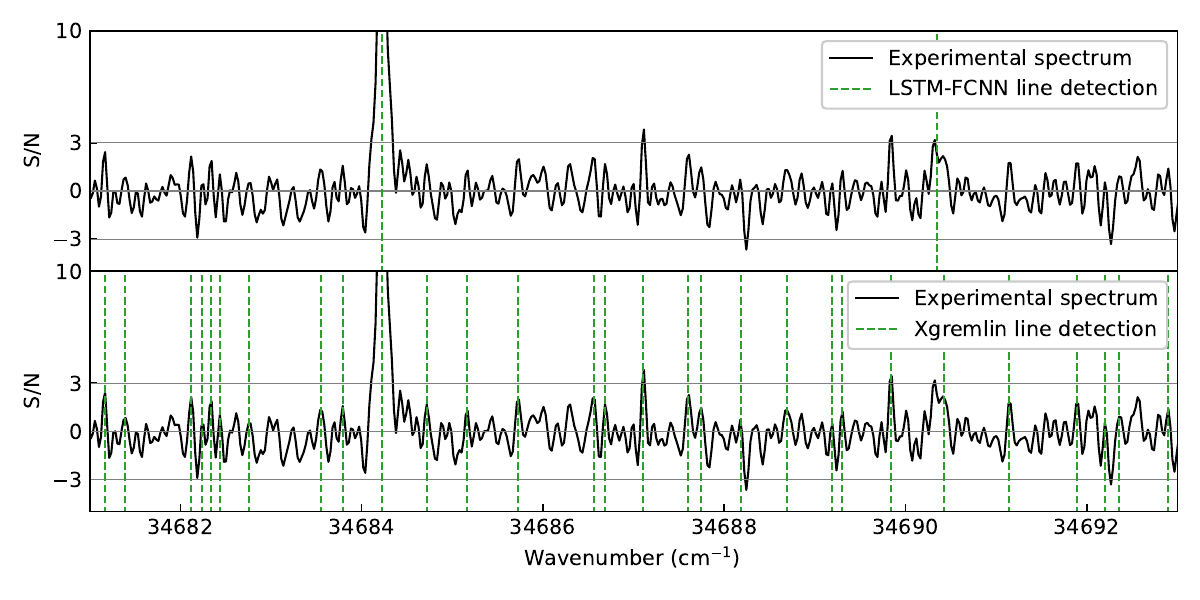}
    \caption{Line detection in the NiHeEH spectrum \cite{clear2022wavelengths} using a LSTM-FCNN model trained using S/N$_{\text{min}}=3$ and evaluated using $p_{\text{th}}=0.5$ (top) and Xgremlin \cite{nave2015xgremlin} using $smin=3$ (bottom).}
    \label{fig7}
\end{figure}

The three 3 S/N lines of Table~\ref{tab:ni2_lines} that were present only in the NN line list were evidently missed by the peak-finding routine of Xgremlin. This was expected because the recommended and typical value for the $smin$ parameter of Xgremlin is 5 \cite{nave2015xgremlin}. When $smin$ was lowered to 3, Xgremlin incorrectly detected a large number of noise spikes as spectral lines. For example, a comparison with LSTM-FCNN line detection with S/N$_{\text{min}}=3$ is shown in Fig.~\ref{fig7} for the vicinity of the 34\,690.3~cm$^{-1}$ line from Table~\ref{tab:ni2_lines}.
The 34\,690.3~cm$^{-1}$ line could not be detected for $smin>3$ in Xgremlin, but setting $smin=3$ introduced too many false line detections and caused a total of 13\,600 lines to be detected within the spectrum. In contrast, the number of lines of this spectrum in the human-made and NN line lists were 561 and 1300, respectively. In Fig.~\ref{fig7}, there are also two spikes above 3 S/N that were undetected by the LSTM-FCNN model. This is not unexpected as they are too narrow compared to all other lines in the spectrum, so the narrow spikes are most likely to be noise, and similar in nature to the few other narrow spikes reaching below negative 3 S/N.

\subsection{Blended lines in the complex Nd lanthanide spectrum}\label{sec:results_blended_lines}
Blends in high-resolution FT atomic spectra are most reliably classified in an atomic energy level analysis, because the analysis of well-resolved spectral lines determine the level energies and hence Ritz wavenumbers that in turn can be used to reliably verify the location of blended spectral lines. For this reason, we investigated whether blended Nd~III lines classified from the energy level analysis of Nd~III \cite{ding2024spectrum1} were correctly identified by the trained LSTM-FCNN model in the high line density Nd-Ar PDL FT spectrum. Results are summarised in Table~\ref{tab:nd3_blends}.
\begin{table}
\caption{\label{tab:nd3_blends}Detection of blended classified Nd~III lines in the Nd-Ar PDL FT spectrum.}
\footnotesize
\begin{tabular}{@{}rcccc}
\br
S/N & $\sigma_{\text{Ritz}}$ & $N_{\text{blend}}$$^{a}$ & $N_{\text{NN}}$$^{b, c}$ & $N_{\text{XG}}$$^{c, d}$ \\
&(cm$^{-1}$)&&&\\
\mr
  5  &  32\,123.642 &  2 &  2(Y) &  2(Y) \\
  6  &  31\,934.185 &  2 &  2(Y) &  2(Y) \\
  5  &  30\,877.685 &  3 &  3(Y) &  3(Y) \\
  5  &  30\,765.741 &  3 &  2(Y) &  1(Y) \\
 53  &  29\,496.112 &  2 &  1(N) &  1(N) \\
 74  &  29\,171.673 &  1 &  1(-) &  1(-) \\
 80  &  28\,745.307 &  2 &  2(Y) &  1(Y) \\
 20  &  28\,259.629 &  2 &  2(Y) &  1(Y) \\
 12  &  27\,255.642 &  2 &  2(Y) &  1(N) \\
527  &  26\,688.545 &  2 &  2(Y) &  1(Y) \\
  9  &  26\,140.225 &  3 &  3(Y) &  2(N) \\
\br
\end{tabular}\\
$^{a}$number of lines in the blend in the human-made line list as proposed in the Nd~III energy level analysis \cite{ding2024spectrum1}; $^{b}$number of lines detected by the trained NN; $^{c}$whether the line of interest is detected at its Ritz wavenumber is indicated in parentheses, `Y' implies yes, `N' implies no, and `-' when not applicable; $^{d}$number of lines detected by Xgremlin with $smin=3$.
\end{table}
\normalsize
The $N_{\text{blend}}$ counts are the number of component lines (as decided by inspection and energy level analysis) fitted in the blend for the line list, where the resulting component line of interest agreed with its expected Ritz wavenumber and relative intensity. The NN line list completely agreed with human analysis for 8 out of the 10 blends with $N_{\text{blend}}>1$. The Nd~III line at 29\,171.673~cm$^{-1}$ with $N_{\text{blend}}=1$ was not an obvious blend to the human and could not be fitted reliably as two lines, it required further consideration of energy level optimisations and analysis of a lower effective temperature HCL spectrum to be classified as a blended feature \cite{ding2024spectrum1}. However, the blended component became visible in the probability curve predicted by the NN model, but below the applied threshold $p_{\text{th}}=0.5$. 
Overall, the LSTM-FCNN model outperformed the conventional peak-finding methods of Xgremlin in analysing blends. For example, the 26\,140.225~cm$^{-1}$ line in the left wing of a strong line (see Fig.~\ref{fig6}) is detected by the trained LSTM-FCNN model, but not by the Xgremlin peak finding routines. Furthermore, setting $smin=3$ for Xgremlin is not ideal as it produced a line list of 13\,500, an extra 6500 and 4100 compared to the human and NN line lists, respectively. Most of these extra lines consisted of spectral features with S/N below 2, similar to the case shown in the bottom plot of Fig.~\ref{fig7}. 

\subsection{Line distortion from instrumental resolution-limited ringing}
The width of the instrumental line profile (FT of Eq.~\ref{eq:instrumental_apodisation} and the cosine bell apodisation) can dominate when the instrumental spectral resolution is insufficient to observe the true emission source line widths. Despite efforts in designing FT spectrometers to have sufficient maximum OPD ($2L$) for their applications, there are situations where instrumental resolution-limited ringing is observed and must be considered, such as many of the FT spectra in the online archives of the Kitt Peak National Solar Observatory (US) \cite{hill1997national, NSO2024}, which are frequently used in atomic spectra analyses (e.g. \cite{clear2022wavelengths,clear2023wavelengths, lawler2022energyzr, lawler2022energyhf}). 

\begin{figure}
    \centering
    \includegraphics[width=\linewidth]{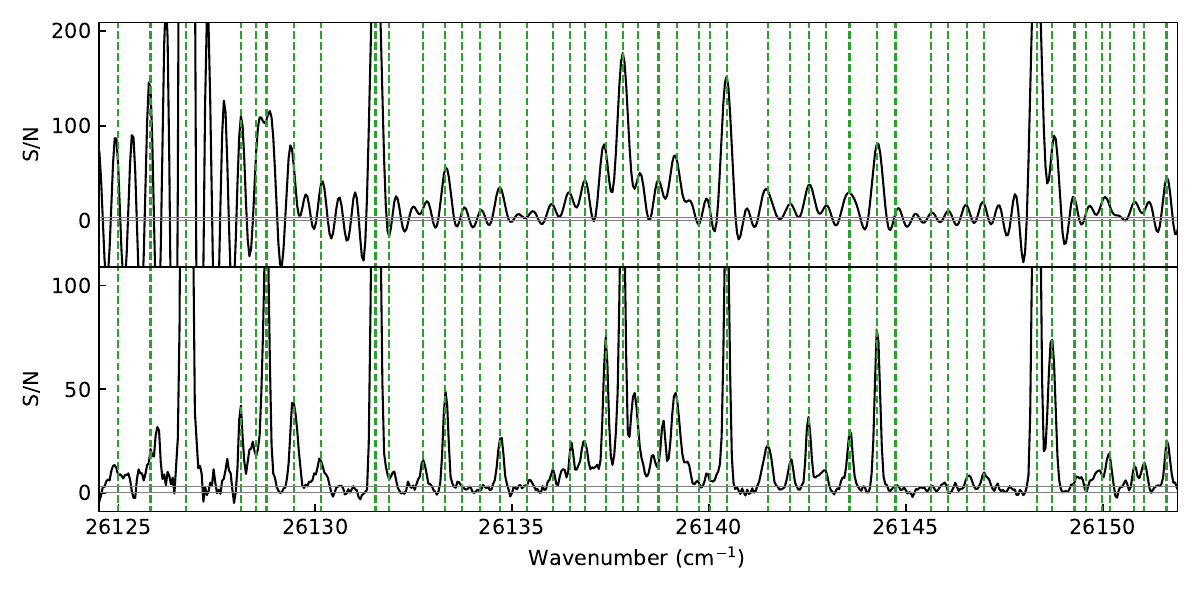}
    \caption{The Nd-Ar FT PDL spectrum with (top) and without (bottom) $4\times$ reduced maximum OPD (instrumental spectral resolution). The solid horizontal lines show zero S/N and the S/N$_{\text{min}}=3$ threshold used for spectrum simulation. The dashed vertical lines show spectral lines detected in the reduced instrumental resolution spectrum (top) by the LSTM-FCNN trained using spectra simulated at the reduced instrumental resolution. The larger S/N range of the top plot is to show the amplitudes of instrumental resolution-limited ringing, the smaller S/N range of the bottom plot is to show weaker line positions.}
    \label{fig8}
\end{figure}

To investigate capabilities of the LSTM-FCNN model in detecting lines distorted or masked by resolution-limited ringing, we artificially introduced resolution-limited ringing to the Nd-Ar PDL FT spectrum by reducing $L$ by a factor of four in the apodisation of its interferogram. We then simulated spectra at this reduced OPD, trained the LSTM-FCNN model, and applied the model to detect lines in the spectrum with reduced OPD. Figure~\ref{fig8} shows an example of the results of this analysis.
Firstly, fewer lines were detected compared to applying and training the model using the full-resolution spectrum, because of the effective reduction in spectral resolution and increase in observed line widths. More importantly, the LSTM-FCNN model has learned the instrumental line profile (FT of Eq.~\ref{eq:instrumental_apodisation} and the cosine bell apodisation) and was able to detect weaker lines masked by the ringing of stronger lines. Furthermore, the LSTM-FCNN model was able to successfully distinguish regions without lines but with ringing of S/N above S/N$_{\text{min}}$ (e.g. 26\,130.8, 26\,141.1, and 26\,147.6~cm$^{-1}$ of Fig,~\ref{fig8}). Several line positions were also detected correctly to not lie on ringing maxima. Obtaining these results are neither feasible nor efficient for a human. The two seemingly undetected lines at 26\,127.7~cm$^{-1}$, lying significantly above S/N$_{\text{min}}$, were potentially results from applying the 20\% threshold in determining whether weaker lines of a blend are detectable in the simulations. However, they are situated in close proximity to a strong 2000 S/N line, which means their detections are also uncertain even in the original full-resolution spectrum due to other significant artefacts arising additionally from phase errors and parts of the instrumental line profile that are not well defined. In the current practice, line detection in instrumental resolution-limited FT spectra is approached by smoothing the spectrum with a kernel to reduce the ringing oscillations in Xgremlin \cite{nave2015xgremlin}. This is far from ideal as any peaks, including ringing features, above the applied S/N threshold would still be detected as lines.

Extra complications arise when determining simulation parameters $G_{\text{w}}$ and $L_{\text{w}}$ for instrumental resolution-limited spectra, because the spectral line fitting must include the instrumental profile and only lines with the highest S/Ns could be fitted reliably. An alternative method could consider estimating the temperature $T$ of the plasma for Eq.~\ref{eq:Gw}. The value of $T$ can be estimated assuming a Boltzmann population distribution for the energy levels and using transition probabilities and relative intensities of the spectral lines. When $L_{\text{w}}$ is significant (e.g. in the IR), one would need to ensure that the LSTM-FCNN model is trained over the possible range of values for $L_{\text{w}}$. 

\section{Conclusions}
Neural networks are concluded to be applicable for line detection in high-resolution FT emission spectra of the open d- and f-shell elements. Improvements over current methods are evident, most notably the increase in line list completeness and the decrease in the number of false line detections will greatly reduce time costs for line list creation. The line position probabilities will also provide a brand new aid to the manual line list adjustment process. We approached the line detection problem with a bidirectional LSTM-FCNN encoder-decoder model trained using simulated spectra of each spectrum of interest. We evaluated the LSTM-FCNN model using spectra of the open 3d-shell Ni (1800--70\,000~cm$^{-1}$) and open 4f-shell Nd (25\,369--32\,485~cm$^{-1}$) elements measured across three different laboratories and a variety of experimental conditions. This approach is shown to be a promising alternative to the peak-finding routines of the widely used Xgremlin program \cite{nave2015xgremlin}. Around 95\% of the spectral lines detected by the trained NN models matched with those in human-made line lists for Ni-He HCL and Nd-Ar PDL spectra, and the discrepancies arose from weak lines, isotope shifts detected as blends, and uncertainties in Voigt profile fitting of frequent blends. Therefore, manual adjustments to the output of the trained NN models are still required. However, this task is expected to be significantly less laborious compared to starting with lines detected by Xgremlin. In our comparisons, the NN model was shown capable of detecting lines around the noise level without introducing as many false detections compared with using Xgremlin. Furthermore, significant fractions of weak and blended lines of the NN output, that were undetected by humans and Xgremlin, were confirmed by the analysis of atomic energy levels. Particularly, the identification of two Ni~II energy levels was enabled by the new lines detected by our LSTM-FCNN models, 3d$^8$$(^3\text{F}_4)6\text{f} \,\,[2]_{3/2}$ at 134\,261.8946 $\pm$ 0.0081 cm$^{-1}$ and 3d$^8$$(^3\text{F}_4)6\text{f} \,\,[1]_{3/2}$ at 134\,249.5264 $\pm$ 0.0054 cm$^{-1}$. These were previously concluded to be unidentifiable due to unobserved weak lines \cite{clear2023wavelengths}.

In addition, the bidirectional LSTM-FCNN encoder-decoder model was shown to be capable of locating lines masked by instrumental resolution-limited ringing. This implied that the LSTM-FCNN model can learn to recognise superimposed complex non-Voigtian line profiles. We expect the model to be able to recognise other line profile distortions such as instrumental artefacts from unconsidered systematic effects, isotope shifts and hyperfine structure in energy levels, and self-absorption due to non-optically thin plasmas in light sources. Accurate simulation of experimental spectra is therefore expected to be key in further improving our approach and reducing manual labour when creating line lists for the FT atomic emission spectra. However, simulating the mentioned effects will be subject to demanding instrumental analyses, atomic structure calculations, and discharge plasma modelling. We also expect new uncertainty estimation methods for wavenumbers and intensities to be developed when applying NNs for line detection. In this work we used NN output line positions as initial parameters for multi-Voigt profile fits of the spectrum, but this approach would not be reliable for lines significantly affected by blending and/or instrumental resolution-limited ringing, which often result in less accurate fitted wavenumbers and intensities compared to their initial guesses, and require further manual adjustments such as fixing fitting parameters. Nevertheless, we hope that this work will begin to enable fellow atomic spectroscopists to invest many more future months into research that are more innovational compared to manually identifying tens of thousands of spectral lines. 

\ack
This work was supported by the Science and Technology Facilities Council (STFC) of the UK and The Bequest of Professor Edward Steers.

\section*{Data and code availability}
The data that support the findings of this study are available from the corresponding author upon reasonable request. The code used for spectrum fitting, spectrum simulation, data processing, neural network training, and line list creation will be available on GitHub after publication or from the corresponding author upon reasonable request.

\section*{References}
\bibliographystyle{iopart-num}
\bibliography{bibliography} 

\end{document}